\documentclass[reprint]{revtex4-1}

\usepackage{graphicx}
\usepackage{dcolumn}
\usepackage{bm}
\usepackage{amsfonts}
\usepackage{amssymb}
\usepackage{amsmath}
\usepackage{float}

\makeatletter

\newcommand{\Rmnum}[1]{\expandafter\@slowromancap\romannumeral #1@}
\makeatother
\begin{document}

\title{Early universe thermostatistics in curved momentum spaces}

\author {M. A. Gorji}\email{m.gorji@stu.umz.ac.ir }
\author{V. Hosseinzadeh}\email{v.hosseinzadeh@stu.umz.ac.ir} \author
{K. Nozari}\email{knozari@umz.ac.ir }\affiliation{Department of Physics,
Faculty of Basic Sciences, University of Mazandaran, P.O. Box
47416-95447, Babolsar, Iran}\author{B.
Vakili}\email{b.vakili@iauctb.ac.ir (Corresponding author)}
\affiliation{Department of Physics, Central Tehran Branch, Islamic
Azad University, Tehran, Iran}

\begin{abstract}
The theories known as doubly special relativity are introduced in
order to take into account an observer-independent length scale and
the speed of light in the framework of special relativity. These
theories can be generally formulated on the de Sitter and also
recently proposed anti-de Sitter momentum spaces. In the context of
these theories, we study the statistical mechanics and to do this,
we consider the natural measure on the corresponding extended phase
space. The invariant measure on the space of distinct microstates is
obtained by restriction of the natural measure of the extended phase
space to the physical phase space through the disintegration
theorem. Having the invariant measure, one can study the statistical
mechanics in an arbitrary ensemble for any doubly special relativity
theory. We use the constructed setup to study the statistical
properties of four doubly special relativity models. Applying the
results to the case of early universe thermodynamics, we show that
one of these models that is defined by the cosmological
coordinatization of anti-de Sitter momentum space, implies a finite
total number of microstates. Therefore, without attribution to any
ensemble density and quite generally, we obtain entropy and internal
energy bounds for the early radiation dominated universe. We find
that while these results cannot be supported by the standard
Friedmann equations, they indeed are in complete agreement with the
nonsingular effective Friedmann equations that arise in the context
of loop quantum cosmology.

\begin{description}
\item[PACS numbers] 04.60.Bc; 05.20.-y
\item[Key Words]
Phenomenological quantum gravity; Statistical mechanics
\end{description}
\end{abstract}
\maketitle
\section{Introduction}
Existence of a minimum measurable length scale, preferably of the
order of the Planck length, is a common feature of the quantum
gravity proposal which is suggested by quantum gravity candidates
such as loop quantum gravity and string theory \cite{LQG,String}.
Although a complete theory of quantum gravity is not yet made, it is
widely believed that purely minimal length effects may be
appreciable when gravity is negligible but the energy scale is very
high \cite{F-QG}. In other words, as general relativity reduces to
the special relativity at the weak gravity limit, it is natural to
expect that a full theory of quantum gravity will be reduced to a
deformed special relativity in which the issues of the existence of
an invariant minimal length and the speed of light are supported
\cite{DSR-RL}. Such a deformed special relativity will reduce to the
standard special relativity at the low energy regime in light of the
correspondence principle. In the absence of a full theory of quantum
gravity, one may do this in reverse: starting from standard special
relativity and deforming it in such a way that in addition to the
speed of light the theory contains a minimal observer-independent
length scale. This is the main idea of the doubly special relativity
(DSR) which was proposed by Amelino-Camelia in Ref. \cite{DSR}.
Lorentz symmetry can be considered as an approximate symmetry that
will be broken at the ultraviolet (UV) regime. However, it is also
possible to construct a DSR theory that preserves Lorentz symmetry
by nonlinear action of the Lorentz group on the momentum space
\cite{DSR-LI}. Indeed, it was realized later that there are many DSR
theories and all of them can be understood as different bases of the
$\kappa$-Poincar\'{e} algebra on the noncommutative Minkowski
spacetime \cite{Glikman-Poincare}. Although there is not a unique
DSR theory, there are some main features which are common between
all DSR theories. For instance, the spacetime structure naturally
turns out to be noncommutative \cite{DSR-NC} in agreement with the
seminal work of Snyder on quantized Lorentz-invariant spacetime
\cite{Snyder}. Interestingly, it is also shown that different DSR
theories can be understood as different coordinate systems on de
Sitter (dS) momentum space
\cite{Glikman-dS1,Glikman-dS2,Curve-M,DSR-RL}. Recently, the anti-de
Sitter (AdS) momentum space was also implemented in the context of
the DSR theories as a complementary to the dS space \cite{AdS,AdS2}.
A maximal momentum or maximal energy corresponding to the universal
observer-independent length scale then arises in these setups.
Therefore, the expressions such as dispersion relations and
invariant measures on the momentum space will be modified which in
turn results modifications to the density of states \cite{DSR-DOS}.
The density of state determines the number of microstates, and
therefore, the thermodynamical properties of the statistical systems
will be significantly affected at the high temperature limit. The
statistical mechanics in the DSR framework was first studied in Ref.
\cite{Glikman}. Thermodynamics of some statistical systems in the
DSR framework are also studied in Ref. \cite{DSR-THR}. It is however
natural to expect that some fundamental aspects of the early
universe thermodynamics may be addressed in the DSR setup. In this
paper, after a brief review on the role of curved dS and AdS
momentum spaces in DSR theories, we introduce four different DSR
models in section \Rmnum{2}. In section \Rmnum{3}, we obtain an
invariant measure on the space of distinct microstates by means of
which one can formulate the statistical mechanics for any DSR model
in any ensemble. In section \Rmnum{4}, we explore the cosmological
applications of the setup when applied to the thermodynamics of the
early radiation dominated universe. Section \Rmnum{5} is devoted to
the summary and conclusions.

\section{Curved Momentum Spaces in DSR Theories}
As we have mentioned above, the DSR theories can be understood as
different coordinate systems on the dS or AdS momentum spaces
\cite{Glikman-dS1,Glikman-dS2,Curve-M,AdS}. However, it should be
noted that while the dS geometry of momentum space can be inspired
by the standard structure of the $\kappa$-Poincar\'{e} Hopf algebra
\cite{Glikman-Poincare,DSR-NC}, such a quantum algebraic structure
is not investigated for the case of the AdS momentum space (see
however Ref. \cite{AdS2}). In fact, taking a minimal
observer-independent length scale into account naturally leads us to
deformed Lorentz transformations \cite{DSR} and curved momentum
spaces \cite{DSR-RL}. The relevance of dS and AdS momentum spaces
with the DSR theories may be easily realized when one notes that the
Minkowski momentum space in the standard special relativity admits
ten isometries and dS and AdS spaces are the only spaces (with
Lorentzian signature in four dimension) that have the same number of
isometries. Furthermore, the constant curvature of these spaces is
consistent with the conjecture of observer independence of the
quantum gravity scale. Thus, implementing these spaces naturally
provides deformed Lorentz transformations that include an
observer-independent quantum gravity scale. Also, since these spaces
are asymptotically equivalent to the Minkowski spacetime
(correspondence principle), the deformed Lorentz transformations
reduce to their standard form in the flat limit (corresponding to
the low energy regime). On the other hand, while the energy and
momentum of a relativistic particle are defined in the usual way in
the standard special relativity, we have freedom to define them in
DSR theories on the curved momentum spaces such that there is no a
clear reason to prefer one basis to another
\cite{Glikman-Poincare,Girelli}. Apart from this feature which shows
the importance of the local properties of the curved momentum spaces
in DSR theories, it is also important to note the global topology of
these spaces. The topologies of dS and AdS are ${\mathbf R}\times
{\mathbf S}^3$ and ${\mathbf S}^1\times{\mathbf R}^3$ respectively.
While a maximal momentum arises by a reasonable identification of
${\mathbf R}$ with the space of energy and ${\mathbf S}^3$ with the
space of momenta in dS momentum space, a maximal energy arises when
one identifies ${\mathbf S}^1$ with the space of energy and
${\mathbf R}^3$ with the space of momenta in AdS momentum space. In
this respect, it seems that dS and AdS momentum spaces will be dual
to each other. However, some interesting features arise in AdS
momentum space which are not predicted by this expected duality, and
therefore, these two spaces are qualitatively different (see Ref.
\cite{AdS}). These standard identifications lead to DSR theories
with isotropic varying speed of light $c=dE/dp$ while other
identifications lead to the nonisotropic varying speed of light (see
Ref. \cite{AdS}). Indeed, the group of symmetries of dS space is
$SO(4,1)$ and the Lorentz symmetry can be preserved by identifying
the Lorentz transformations with the six elements of the subgroup
$SO(3,1)$ and the four remaining generators with the positions
\cite{Glikman-dS1}. In the same manner, the symmetry group of AdS
space is $SO(3,2)$, and the Lorentz invariance can be preserved
through the identification of the Lorentz transformations with the
subgroup $SO(3,1)$ of $SO(3,2)$. Interestingly, the commutation
relations between the positions belonging to the two quotients of
two algebras $so(4,1)/so(3,1)$ in dS and $so(3,2)/so(3,1)$ in AdS
turn out to be noncommutative \cite{Glikman-dS1,Glikman-dS2}. This
feature is general for any DSR theory on different coordinate
systems on dS or AdS momentum spaces. The space of four-momenta then
will be the quotient spaces $SO(4,1)/SO(3,1)$ and $SO(3,2)/SO(3,1)$
in the case of dS and AdS momentum spaces respectively. Clearly, the
corresponding extended eight-dimensional phase space is
noncommutative with topology ${\mathbf R}^4\times{\mathbf{dS}}$
\cite{Glikman-dS2} and ${\mathbf R}^4\times {\mathbf{AdS}}$. At the
flat low energy limit, the minimal observer-independent effects
become negligible and both dS and AdS reduce to the Minkowski space
with the standard phase space with ${\mathbf R}^4\times{\mathbf
R}^4$ topology. It is also interesting to note that while gravity
can be understood as the curvature of the spacetime sector in a
general theory of relativity, a minimal observer-independent length
scale, as a universal UV cutoff, can be understood as a constant
curvature of the momentum sector of the extended phase space in DSR
theories, motivated by the effects of invariant quantum gravity
scale on general relativity, known as gravity's Rainbow, also can be
considered \cite{Rainbow}.

The curved four-momentum spaces in DSR theories then can be
realized from the four-dimensional hypersurfaces \cite{AdS}
\begin{equation}\label{hypersurface-4}
-P_0^2+P_1^2+P_2^2+P_3^2\pm{P_4^2}=\pm{l^{-2}}\,,
\end{equation}
which are embedded in five-dimensional flat spaces with signatures
$(-,+,+,+,+)$ and $(-,-,+,+,+)$ for dS and AdS cases, respectively
\cite{unit}. In relation (\ref{hypersurface-4}), $P_A$ terms with
$A=0,..,4$ are embedding coordinates and $l$, with dimension of
length, is the radius which signals an observer-independent length
scale. In order to ensure that the quantum gravity effects become
important just at the very high energy regime, the invariant length
scale $l$ is usually assumed to be of the order of the Planck length
$l=\beta_0\,l_{_{\rm Pl}}$ where $\beta_0={\mathcal O}(1)$ should be
fixed by the experiments \cite{QG-EXP}. The corresponding line
elements then will be
\begin{equation}\label{metric-5}
ds^2=-dP_0^2+dP_1^2+dP_2^2+dP_3^2\pm{dP_4^2}\,.
\end{equation}
In relations (\ref{hypersurface-4}) and (\ref{metric-5}), the $(+)$
and $(-)$ signs denote the dS and AdS spaces respectively. One then
can consider particular coordinate system on both dS and AdS
momentum spaces by fixing the embedding coordinates $P_A$ in terms
of physical energy and momenta. As a common way one may solve the
constraint relation (\ref{hypersurface-4}) for $P_4$ and then
substitute the result into relation (\ref{metric-5}) which gives the
metric of the four-dimensional curved momentum space. In this way,
all the well-known DSR theories can be realized by a suitable fixing
of the embedding coordinates $P_A$. Interestingly, the Snyder
algebra \cite{Snyder} can be derived from this setup \cite{Girelli},
and therefore, it is nothing but a particular DSR theory. It is also
possible to consider a deformed relativistic algebra in which the
embedding coordinate $P_4$ has not been removed and is present in
the resultant associated four-dimensional algebra. Such an algebra,
for instance, is investigated in Ref. \cite{Mendes} in the context
of the stability theory of the Lie algebras in which $P_4$ plays the
role of nontrivial center of the resultant deformed algebra. In
comparison with the Snyder and other DSR algebras such as bi-cross
product algebra, this stable algebra cannot be considered as a
closed algebra in four dimensions \cite{Girelli}.

Among all the possible coordinate systems on dS and AdS momentum
spaces, the natural coordinate system on dS momentum space is
inspired by the bi-cross product basis of $\kappa$-Poincar\'{e}
algebra which is known as the cosmological coordinates since it
corresponds to the cosmological rendition of dS space in position
space. This DSR theory inspired the deformed Lorentz transformation
such that the Lorentz symmetry is preserved. On the other hand, its
counterpart on AdS momentum space, {\it i. e.} the DSR theory
defined by cosmological coordinates on AdS momentum space, breaks
the Lorentz symmetry. The Lorentz invariant DSR theory is then found
in static coordinatization of AdS momentum space \cite{AdS}. Also,
the static coordinatization of dS momentum space is investigated,
which breaks the Lorentz invariance. Although the DSR theories are
different from each other, as a candidate for the flat limit of
ultimate quantum gravity theory, all of them are possible, and there
is not a clear physical reason to prefer one over the other. In the
next subsections, we therefore review the results of Ref. \cite{AdS}
for dS and AdS momentum spaces in both of the cosmological and
static coordinates. Our task is to generally formulate the
statistical mechanics in DSR theories defined on curved momentum
spaces and then compare the different DSR theories (that arise from
different coordinatization on momentum spaces) from the
thermostatistical point of view.

\subsection{de Sitter momentum space}
\subsubsection{Cosmological coordinate (dS-Cosm model)}
The relations between the induced physical energy and momenta
$(E,p_i)$ and the embedding coordinates in the cosmological
coordinate system are defined as \cite{AdS}
\begin{align}\label{ds-embed}
&P_0(E,\vec{p})=\frac{1}{l}\sinh(lE)+\frac{lp^2}{2}\exp(lE),\nonumber\\
&P_i(E,\vec{p})=-p_i\exp(lE),\\
&P_4(E,\vec{p})=-\frac{1}{l}\cosh(lE)+\frac{lp^2}{2}\exp(lE),\nonumber
\end{align}
where $p=|\vec{p}|=\sqrt{\delta^{ij}p_ip_j}$ with $i,j=1,2,3$.
Rewriting line element (\ref{metric-5}) with a $(+)$ sign
(corresponds to dS space) in terms of the physical energy and
momenta $(E,p_i)$ defined by the above relations and then removing
$P_4$ by means of constraint (\ref{hypersurface-4}) (again with a
$(+)$ sign), the line element of dS momentum space works out to be
\begin{equation}\label{metric-ds},
ds^2=-dE^2+\exp(2lE)\sum_{i=1}^{3}dp_i^2\,.
\end{equation}
The above line element gives the invariant integration measure
\cite{convention}
\begin{equation}\label{ds-d4p}
\frac{d\mu(E,\vec{p})}{4\pi}=\exp(3lE)dEp^2dp\,,
\end{equation}
on the momentum space. The corresponding deformed mass-shell
condition is determined by demanding $P_4$ to be constant in
(\ref{hypersurface-4}) as $-P_0^2+{\vec{P}}^2=l^{-2}-P_4^2=m^2$
with $P_0>0$ and $P_4<0$. For the massless case $m=0$, with
which we are interested in this paper, it gives
\begin{equation}\label{mass-shell-ds0}
{\mathcal C}\left(1+\frac{l^2{\mathcal C}}{4}\right)=0\,,
\end{equation}
where
\begin{equation}\label{mass-shell-ds}
{\mathcal C}=-\frac{4}{l^2}\sinh^2(lE/2)+p^2\exp(lE)\,.
\end{equation}
Solving the above constraint gives the modified dispersion relation
$E=-l^{-1}\ln(1-lp)$. This dispersion relation also shows that there
is a maximal momentum as $p\leq{1/l}$ which is the consequence of
compact ${\mathbf S}^3$ topology of the space of momenta \cite{LT}.
The energy $E$, however, can take any positive value as
$E\in[0,\infty)$. In the flat low energy limit $lE\propto{E/E_{_{\rm
Pl}}}\ll1$, the line element (\ref{metric-ds}) reduces to the flat
case $ds^2\approx-dE^2+\sum_{i=1}^{3}dp_i^2$, the invariant measure
(\ref{ds-d4p}) reduces to the standard well-known measure
$d\mu(E,\vec{p})={4\pi}dEp^2dp$, and the deformed mass-shell
condition (\ref{mass-shell-ds}) also leads to the standard
Einsteinian dispersion relation ${\mathcal C}=-E^2+p^2$ for the
massless particles with $E,p\in[0,\infty)$. The modified dispersion
relation (\ref{mass-shell-ds}) immediately leads to the varying
speed of light $c=\frac{dE}{dp}=\exp(-lE) =(1-lp)^{-1}$ (see also
Ref. \cite{AdS}).
\subsubsection{Static coordinate (dS-Stat model)}
We do not repeat all the calculations for the DSR theory defined by
the static coordinatization of dS momentum space and only review the
main results we deal with throughout this paper (see Ref. \cite{AdS}
for more details).

The line element associated with the static coordinate system
defined on dS momentum space is given by
\begin{equation}\label{metric-ds-s}
ds^2=-(1-l^2p^2)dE^2+\frac{dp^2}{1-l^2p^2}+p^2d\Omega^2\,,
\end{equation}
where clearly $d\Omega^2=d\theta^2+\sin^2\theta{d\varphi^2}$ is the
metric of the two-sphere with unit radius. The corresponding
invariant integration measure is
\begin{equation}\label{ds-d4p-s}
\frac{d\mu(E,\vec{p})}{4\pi}=dEp^2dp\,,
\end{equation}
which remains unchanged. For the massless particles, the Casimir is
given by
\begin{equation}\label{mass-shell-ds-s}
{\mathcal C}=-\frac{1}{l^2}\sinh^2(lE)(1-l^2p^2)+p^2=0\,,
\end{equation}
which gives the modified dispersion relation $E=
l^{-1}\tanh^{-1}(lp)$ with $p\leq{1/l}$ and $E\in[0,\infty)$. In the
low energy limit $lE\propto{E/E_{_{\rm Pl}}}\ll1$, the line element
(\ref{metric-ds-s}) correctly reduces to the flat Minkowski one and
the deformed mass-shell condition (\ref{mass-shell-ds-s}) leads to
the usual dispersion relation ${\mathcal C}=-E^2+p^2$ for the
massless particles with $E,p\in[0,\infty)$. The modified dispersion
relation (\ref{mass-shell-ds-s}) also leads to the varying speed of
light as $c=\frac{dE}{dp} =(1-l^2p^2)^{-1}$ in this chart.

\subsection{Anti-de Sitter momentum space}
\subsubsection{Cosmological coordinate (AdS-Cosm model)}
For the case of the cosmological coordinate system on AdS momentum
space, the relations between induced physical energy and momenta
$(E,p_i)$ and the embedding coordinate are defined as follows
\cite{AdS}:
\begin{align}\label{ads-embed}
&P_0(E,\vec{p})=\frac{1}{l}\sin(lE),\nonumber\\
&P_i(E,\vec{p})=p_i\cos(lE),\\
&P_4(E,\vec{p})=\frac{1}{l}\cos(lE)\sqrt{1+l^2p^2}.\nonumber
\end{align}
Similar to the case of dS momentum space, substituting from the
above relations into the relation (\ref{metric-5}) and then removing
$P_4$ from the $(-)$ sign of constraint (\ref{hypersurface-4}), the
line element of AdS momentum space in terms of physical energy and
momenta $(E,p_i)$ turns out to be
\begin{equation}\label{metric-ads}
ds^2=-dE^2+\cos^2(lE)\left(\frac{dp^2}{1+l^2p^2}+p^2d\Omega^2
\right)\,.
\end{equation}
The invariant measure on the momentum space then will be
\begin{equation}\label{ads-d4p}
\frac{d\mu(E,\vec{p})}{4\pi}=\cos^3(lE)dE\frac{p^2dp}{
\sqrt{1+l^2p^2}}\,.
\end{equation}
The associated mass-shell condition $-P_0^2+{\vec{P}}^2=
l^{-2}-P_4^2=m^2$ then leads to
\begin{equation}\label{mass-shell-ads}
{\mathcal C}=-\frac{1}{l^2}\sin^2(lE)+p^2\cos^2(lE)=0\,,
\end{equation}
for the massless case $m=0$. By solving the above relation one can
easily find the modified dispersion relation $E=l^{-1}\tan^{-1
}(lp)$ which clearly implies a maximal energy $E\leq+\pi/2l$, and
thus in relation (\ref{ads-embed}) we have $0\leq{P_0} \leq+1/l$.
Relations (\ref{metric-ads}), (\ref{ads-d4p}), and
(\ref{mass-shell-ads}) reduce to their standard counterparts in the
flat low energy limit $lE\propto{E/ E_{_{\rm Pl}}}\ll1$. The
modified dispersion relation (\ref{mass-shell-ads}) also implies the
variation of the speed of light at high energy regime as
$c=\frac{dE}{dp}= \cos^2(lE)=(1+l^2p^2)^{-1}$ \cite{AdS}.
\subsubsection{Static coordinate (AdS-Stat model)}
In static coordinatization of AdS momentum space, the line
element takes the following form
\begin{equation}\label{metric-ads-s}
ds^2=-(1+l^2p^2)dE^2+\frac{dp^2}{1+l^2p^2}+p^2d\Omega\,,
\end{equation}
and therefore the integration measure remains unchanged as
\begin{equation}\label{ads-d4p-s}
\frac{d\mu(E,\vec{p})}{4\pi}=dEp^2dp\,.
\end{equation}
The Casimir invariant is given by
\begin{equation}\label{mass-shell-ads-s}
{\mathcal C}=-\frac{1}{l^2}\sin^2(lE)(1+l^2p^2)+p^2=0\,,
\end{equation}
for the massless case which implies the modified dispersion
relation $E=l^{-1}\tan^{-1}(lp)$ with $p\in[0,\infty)$ and
$E\leq{\pi/2l}$. This modified dispersion relation implies
the varying speed of light as $c=\frac{dE}{dp} =(1+l^2p^2
)^{-1}$.
\section{Statistical Mechanics: Invariant Measure}
In this section, we generally formulate the statistical mechanics
for DSR theories. The standard statistical mechanics is based on the
invariant measure (density of states) of the physical phase space of
a nonrelativistic system which is a six-dimensional symplectic
manifold (for a system consisting of one particle) and determines
the number of microstates at the semiclassical regime. In the case
of relativistically formulated theories like DSR theories, we deal
with an invariant measure on an extended eight-dimensional phase
space. In order to formulate the statistical mechanics for such
systems, we should be able to count the number of distinct
accessible microstates of the system by finding the corresponding
appropriate measure on its physical phase space \cite{rovelli}. Some
attempts have been made in this direction to study the different
statistical systems in the context of DSR theories (see Refs.
\cite{Glikman,DSR-THR}). But, here, we would like to generally
formulate the statistical mechanics of DSR theories in a more
systematic way. The approach we introduce allows one to study the
thermodynamical properties of statistical systems in any ensemble
for any DSR theory.

To find the invariant measure on the six-dimensional physical phase
space, we start with a natural measure on the eight-dimensional
extended phase space $\Gamma_X \equiv(t,\vec{x};E,\vec{p})$ that is
given by \cite{GHS}
\begin{eqnarray}\label{measure-ex}
\mu_X=\int d\mu_{X}=\int d\mu(t,\vec{x})d\mu(E,\vec{p})\,,
\end{eqnarray}
where $d\mu(t,\vec{x})$ and $d\mu(E,\vec{p})$ are the standard
invariant volume elements on the spacetime and momentum sector of
$\Gamma_X$ which are defined by the metrics on these spaces. The
metric on the spacetime sector is flat since DSR theories are the
flat limit of the ultimate quantum gravity theory (in the limit
where gravity is negligible) \cite{F-QG}, and therefore, the
curvature of the spacetime sector will be zero \cite{IR-TSR}. It is,
however, important to note that while the metric is flat, it is not
defined on the standard (commutative) Minkowski spacetime. But, it
is indeed defined on a noncommutative $\kappa$-Minkowski spacetime
($\kappa\sim{l^{ -1}}$ in our notation) dual to the corresponding
curved momentum space \cite{DSR-NC}. For many-particle systems that
one usually considers in field theory and statistical mechanics,
there is not an appropriate (well-defined) measure on noncommutative
spacetime which respects all the desired symmetries. More precisely,
the standard Lebesgue measure $d\mu(t,\vec{x})=d^4x=dt\,d^3x$
respects the $\kappa$-Poincar\'{e} symmetries while it evidently
cannot support the cyclicity of the action functional (see Ref.
\cite{Agostini} for more details). Trying to recover the cyclicity
of the action functional, one, however, should renounce the
$\kappa$-Poincar\'{e} invariance of the theory and also the
correspondence principle such that the standard commutative
Minkowski spacetime would not be obtained from the corresponding
$\kappa$-Minkowski spacetime in the low energy limit
$\kappa\rightarrow\infty$ (or equivalently $l \rightarrow\,0$). The
problem is not yet definitively answered. In this respect, we
consider the standard Lebesgue measure
$d\mu(t,\vec{x})=d^4x=dt\,d^3x$ for the configuration space of
$\Gamma_X$ which respects both the $\kappa$-Minkowski spacetime
structure and correspondence principle. For the momentum sector, the
measure is completely defined by the metric on curved dS or AdS
momentum spaces as $d\mu(E,\vec{p})=4\pi\sqrt{-g}d^4p$ where $g$ is
the determinant of the metric of the associated curved momentum
space.

Measure (\ref{measure-ex}), however, is not restricted to the
dispersion relation while the permitted and physically relevant
microstates are those that are laid on the constraint given by the
dispersion relation. Note also that the pullback of $d\mu_{X}$ to
the constraint surface (defined by the corresponding dispersion
relation) vanishes since $d\mu_X$ is an eight-form and the
constraint is defined on a seven-dimensional space. The useful tool
is the so-called disintegration theorem that lets us to restrict the
measure (\ref{measure-ex}) of the eight-dimensional phase space
$\Gamma_X$ to the seven-dimensional space (defined by the associated
dispersion relation) as (see also Appendix A of Ref. \cite{rovelli})
\begin{eqnarray}\label{measure-c}
\mu_\mathcal{C}=\int \delta(\mathcal{C})\,d\mu_{X}\,,
\end{eqnarray}
where ${\mathcal C}=0$ is clearly the constraint equation. One can
check that $\mu_\mathcal{C}$ has all the properties of a measure.
The points on the constraint, however, are not totally distinct
microstates. Indeed, the constraint is subdivided into the
equivalent classes of microstates which are linked by the time
evolution that is generated by the constraint itself (orbits). They
are physically equivalent since the time evolution induced by the
constraint is nothing but a gauge transformation in the
relativistically formulated theories such as DSR theories. Taking
the coordinate $t$ to be time, we can parametrize microstates in
each set by $t$. Then, to obtain the space of the distinct
microstates, we must consider only one microstate of each equivalent
class (gauge fixing). This can be done by choosing the slice (that
is a six-dimensional manifold) of $t=t_0$ on the constraint that is
appropriately intersecting with orbits. This space provides the
physical phase space of the system and one can find a measure on it
by $\mu_\mathcal{C}$ and the constraint $\delta(t-t_0)$ through the
disintegration theorem. Using again the disintegration theorem to
fix the gauge, the appropriate measure on the physical phase space
then turns out to be
\begin{align}\label{measure-p}
\mu_{p}&=\int d\mu_{p}=\int\delta(\mathcal{C})\delta(t-t_0)
\,d\mu_{X}\\&=\int\delta(\mathcal{C})\delta(t-t_0)\,d\mu(t,
\vec{x})d\mu(E,\vec{p})\,\nonumber.
\end{align}
This is a natural measure on the space of distinct and physically
relevant microstates for a statistical system. Measure
(\ref{measure-p}) is indeed nothing but the density of states when
one selects particular ensemble density such as Dirac delta
function, Boltzmann factor, Bose-Einstein, or Fermi-Dirac ensemble
densities for microcanonical, canonical, Bose-Einstein, and
Fermi-Dirac statistics, respectively. Having measure
(\ref{measure-p}) at hand, we are then adequately equipped to
generally formulate the statistical mechanics for DSR theories in
any ensemble.

\section{Early Universe Thermodynamics}
One of the most interesting features of the presented setup is its
application to the thermodynamics of the early radiation dominated
universe. One then should study the statistical mechanics of the
effectively massless particles (bosons and fermions) which
contribute to the energy content of the radiation dominated
universe. Measure (\ref{measure-p}), however, does not an analytical
solution for the Bose-Einstein and Fermi-Dirac ensemble densities
for four models that are introduced in this paper. Nevertheless, we
will introduce an approach with which one can realize general
features of any DSR theory, which is important for early universe
thermodynamics, without attribution to any ensemble density.

\subsection{Total number of microstates}
To do so, we consider the total number of microstates for a particle
in a DSR framework. In statistical mechanics, the density of states
determines the number of accessible microstates for the system under
consideration. The density of states is determined by measure
(\ref{measure-p}) when one fixes a particular ensemble density. For
example, one should consider the well-known Bose-Einstein and
Fermi-Dirac ensemble densities to study the statistical mechanics of
the early radiation dominated universe. But, what do all the
ensemble densities do in statistical mechanics formalism? They
indeed define the probability distribution over the set of all
microstates by restricting the system to the subset of accessible
microstates from the infinite set of all microstates that the system
can potentially access. The number of total microstates is
determined by measure (\ref{measure-p}) without attribution to any
ensemble density. To be more precise, the particles in the early
Universe (such as photons and electrons) are nonlocalized, and
therefore, the spacetime part of the measure (\ref{measure-p})
simply reduces to the physical volume $V$ (in which the particles
are confined) as
\begin{eqnarray}\label{NoM}
\Omega=\frac{V}{h^3}\,\int\delta(\mathcal{C})\,
d\mu(E,\vec{p})\,,
\end{eqnarray}
in which, to take into account the Heisenberg uncertainty principle
in semiclassical statistical mechanics, we have divided the measure
(\ref{measure-p}) by $h^3$, with $h$ being the Planck constant. Note
that although $h=2\pi$ in our unites since $\hbar=1$ \cite{unit}, we
explicitly work with $h$ rather than $2\pi$ to show its significance
in the determination of the number of microstates. For the standard
early universe thermodynamics, the bosons and fermions obey the
usual Einsteinian dispersion relation $E=p$ with nondeformed
measure, and the total number of microstates (\ref{NoM}) is
diverging as
\begin{equation}\label{NoM-SR}
\Omega(V)=\frac{4{\pi}V}{h^3}\,\int_0^{\infty}E^2\,
dE\rightarrow\infty\,.
\end{equation}
It is important to note that in the standard statistical mechanics
of the early Universe, the Bose-Einstein and Fermi-Dirac ensemble
densities select a finite subset of microstates from the infinite
total number of microstates (\ref{NoM-SR}) as the accessible
microstates for the system. But, however, the system can access more
and more microstates by increasing the energy. The reason for which
we have considered the total number of microstates will become clear
when one is interested in DSR theories which predict an upper bound
for the total energy of the system. Let us calculate (\ref{NoM}) for
the four different DSR models that are introduced in this paper.
Using the results of section \Rmnum{2} in relation (\ref{NoM}),
leads to the following results
\begin{eqnarray}\label{NoM-DSR}
\Omega(V,l)=\left\{
\begin{array}{llll}
\frac{{16\pi}V}{h^3l^2}\int_0^{\infty}e^{2lE}\sinh^2({lE}/{2}
)dE\rightarrow\infty, & \mbox{dS-Cosm}\vspace{.5cm}\\
\frac{{4\pi}V}{h^3l^2}\int_0^{\infty}\tanh^2(lE)dE\rightarrow
\infty, & \mbox{dS-Stat}\vspace{.5cm}\\
\frac{{\pi}V}{h^3l^2}\int_0^{\frac{\pi}{2l}}\sin^2(2lE)dE=
\frac{\pi^2V}{4h^3l^3}, & \mbox{AdS-Cosm}\vspace{.5cm}\\
\frac{{4\pi}V}{h^3l^2}\int_0^{\frac{\pi}{2l}}\tan^2(lE)dE
\rightarrow\infty, & \mbox{AdS-Stat}
\end{array}
\right.
\end{eqnarray}
The above results show that the number of total microstates for a
particle is infinite for dS-Cosm, dS-Stat, and AdS-Stat models while
it is {\it finite} for the case of the AdS-Cosm model. Let us
elaborate more on the results (\ref{NoM-DSR}). By increasing the
kinematical energy $E$, a statistical system can access more and
more microstates. For DSR theories with maximal energy such as
AdS-Cosm and AdS-Stat, this process cannot infinitely continue since
there is an upper bound $E\in[0,\pi/2l)$, while in standard special
relativity we have ($E\in[0, \infty$)). The compact $S^1$ topology
for the DSR theories that are defined on AdS momentum space, makes
the total volume of the energy space always finite \cite{LT}.
Measure (\ref{NoM}) (or (\ref{measure-p})) is however defined on the
whole of the momentum space including the space of momenta $\vec{p}$
that is not compact for the AdS case. Therefore, having just compact
energy space cannot make the total number of microstates (\ref{NoM})
finite. The space of momenta $\vec{p}$ affects the total number of
microstates (\ref{NoM}) from constraint ${\mathcal C}$. In this
respect, apart from the compact topology of the energy space which
is a necessary condition to have a finite number of microstates, we
should also explore another enough condition which would explain how
the number of total microstates is finite in the AdS-Cosm model
while it is infinite for the AdS-Stat case. The key is indeed the
dimensional reduction at the UV regime which leads to the reduction
of the number of microstates in this regime. This is the common
feature of almost all quantum gravity candidates \cite{DR}, and the
DSR models also predict dimensional reduction at the UV regime. More
precisely, by parametrizing the constraint by integer $\gamma $ as
${\mathcal C}(1+l^{2\gamma}{\mathcal C}^{ \gamma})$, one can realize
dynamical dimensional reduction for the Hausdorff dimension of the
momentum space. Following Ref. \cite{DSR-DR}, one also identifies
the Hausdorff dimension of the momentum space with the spectral
dimension of the spacetime sector and then interprets the Hausdorff
dimensional reduction of the momentum space as the spectral
dimension reduction in the spacetime sector \cite{DR-NC}. For the
four models which we have considered in this paper with the standard
Hausdorff dimension of momentum space equal to $4$ in the IR regime,
the Hausdorff dimension at the UV regime runs as
\begin{eqnarray}\label{UV-H-dimension}
d_H(4,\gamma)=\left\{
\begin{array}{llll}
\frac{6}{1+\gamma}, & \hspace{.5cm}\mbox{dS-Cosm}\vspace{.5cm}\\
\frac{3}{1+\gamma}, & \hspace{.5cm}\mbox{dS-Stat}\vspace{.5cm}\\
\frac{3}{1+\gamma}, & \hspace{.5cm}\mbox{AdS-Cosm}\vspace{.5cm}\\
\frac{4}{1+\gamma}, & \hspace{.5cm}\mbox{AdS-Stat}
\end{array}
\right.
\end{eqnarray}
From the results of section \Rmnum{2}, it is clear that $\gamma=1$
for the dS-Cosm model and $\gamma=0$ for the three other models.
According to (\ref{UV-H-dimension}), the dS-Cosm, dS-Stat and
AdS-Cosm models imply the dynamical dimensional reduction by $1$ at
the UV regime while the AdS-Stat model does not. Note that the
AdS-Cosm model is the only model that has both the necessary and
enough conditions: compact energy space and also dimensional
reduction in the UV regime. We conjecture that these conditions are
sufficient to have a finite total number of microstates for the
statistical systems. The results (\ref{NoM-DSR}) and
(\ref{UV-H-dimension}) show that the DSR theories have very
different behaviors at the UV regime from the thermostatistical and
kinematical point of views. Our thermostatistical consideration has,
however, an advantage that it selects the AdS-Cosm model to be more
relevant in comparison to the other models. This is because of the
finite total number of microstates that emerged in statistical
consideration of this model. Existence of a finite total number of
microstates immediately leads to an entropy bound for the system
under consideration which is also a common feature of quantum
gravitational systems such as black holes \cite{BH-Entropy}.
\subsection{Entropy and energy density bounds}
To obtain the entropy bound, we note that in the AdS-Cosm model with
a finite total number of microstates (\ref{NoM-DSR}) for one
particle, the total number of microstates for the system consisting
of $N$ such particles will be $\Omega_N=\Omega^N/N!$ where the Gibbs
factor is also considered since the particles are indistinguishable.
The associated maximum entropy $S_{\max}=\ln\Omega_N$ then takes the
following form
\begin{equation}\label{Entropy-Bound}
\frac{S_{\max}}{N}=\ln\left(\frac{V}{h^3l^3}
\right)+\ln\left(\frac{\pi^2}{4N}\right)+1\,,
\end{equation}
in which we have used (\ref{NoM-DSR}) and also the Stirling’s
approximation $\ln{N!}=N\ln{ N}-N$. Taking the fact that $hl\sim{
l_{_{\rm Pl}}}$ into account in relations (\ref{NoM-DSR}) and
(\ref{Entropy-Bound}), one can see that the total number of
microstates for the universe is precisely determined by the factor
$V/l_{_{\rm Pl}}^3$. This result shows that the fundamental volume
of microstates for the quantum gravitational statistical system will
be proportional to $l_{_{\rm Pl}}^3$ with which the physical volume
$V$ is quantized. We note however that the fundamental volume of
microstates in the standard statistical system (for a particle) is
$h^3$ with which the phase space volume is quantized. In some senses
this feature is similar to the case of Bekenstein-Hawking entropy of
black holes where the number of microstates is determined by the
factor $A/l_{_{\rm Pl}}^2 $ with $A$ being the horizon area of the
black hole \cite{BH-Entropy,PH-EB}.

On the other hand, the internal energy $U$ is the average of the
kinematical energy $E$, and the total potentially accessible
internal energy for the statistical system consisting of $N$
particles in this setup then will be
\begin{eqnarray}\label{Internal-Energy}
U_{\rm tot}=N\times\left(\frac{\int{E}\,
d\mu_p}{\int{d}\mu_p}\right).
\end{eqnarray}
Substituting the Einsteinian dispersion relation into the above
relation, one realizes that this relation is diverging in the
framework of standard special relativity. From the statistical point
of view, this is because of the fact that the system can access more
and more microstates by increasing the kinematical energy $E$. This
possibility in standard special relativity leads to the well-known
cosmological feature that there is not an upper bound for the energy
density of the early radiation dominated universe (the standard
Stefan-Boltzmann law). Calculating relation (\ref{Internal-Energy})
for the four presented DSR models, we deduce that it is diverging
for all the models except the AdS-Cosm model. In this case, relation
(\ref{Internal-Energy}) converges to
\begin{eqnarray}\label{Internal-Energy-Bound}
\frac{U_{\max}}{N}=\frac{4l}{\pi}\int_0^{\frac{
\pi}{2l}}\sin^2(2lE)EdE=\frac{\pi}{4l}\,.
\end{eqnarray}
The above result shows that the existence of the upper bound
$E\leq{\pi/2l}$ together with the dimensional reduction at the UV
regime (which drastically reduces the number of microstates) leads
to the nontrivial upper bound (\ref{Internal-Energy-Bound}) for the
internal energy as $U\leq{U_{\max}} \sim{E_{_{\rm Pl}}}$. This bound
together with the entropy bound (\ref{Entropy-Bound}) leads to an
upper bound for the energy density of the early radiation dominated
universe as $\rho\leq\rho_{\max}$ with
$\rho_{\max}=\frac{U_{\max}}{V}=\frac{ (U_{\max}/N)}{(V/N)}$ which
after substituting $(V/N)$ from (\ref{Entropy-Bound}) and some
manipulations is given by
\begin{equation}\label{rho-bound}
\rho_{\max}=\frac{e\pi^3}{16h^3l^4}\,
\exp\left({-\frac{S_{\max}}{N}}\right).
\end{equation}
The above relation shows that $\rho\leq
\rho_{\max}\sim{T_{_{\rm Pl}}^4}$.

\section{Cosmology: DSR versus LQC}
What is the cosmological implication of bounds (\ref{Entropy-Bound})
and (\ref{rho-bound})? Consider the standard Friedmann equation
\begin{equation}\label{Friedmann}
H^2=\frac{8{\pi}G}{3}\rho\,,
\end{equation}
where $H=\dot{a}/a$ is the Hubble parameter with $a$ being the scale
factor. At first glance, one can deduce that the existence of the
upper bound (\ref{rho-bound}) for the energy density implies an
upper bound for the Hubble parameter through the Friedmann equation
(\ref{Friedmann}). To be more precise, one should note that the
geometric part of the Friedmann equation is completely determined by
the classical Einstein's equations which do not predict any upper
bound for the Hubble parameter. Indeed, it is the standard
statistical mechanics that is consistent with the standard classical
Einstein's equations such that both the Hubble parameter and energy
density of the radiation dominated universe diverge at a big bang
leading to the so-called big bang singularity problem. We however
notice that the upper bound that arises for the energy density of
the radiation dominated universe in our setup is due to the quantum
gravitational (minimal length) effects. Thus, the inconsistency
between the right- and left-hand sides of the Friedmann equation
arises when one applies the quantum gravitational effects for the
matter content while considering the geometric part to be purely
classic. We should therefore explore quantum gravitational effects
for the geometric part which support the energy density bound
(\ref{rho-bound}) and also entropy bound (\ref{Entropy-Bound}) that
we have obtained in the DSR framework. Very interestingly, the
modified Friedmann equations that are suggested by loop quantum
cosmology (LQC) \cite{LQC} predict an upper bound for the energy
density. The modified Friedmann equation for the flat early
radiation dominated universe is given by \cite{Ashtekar}
\begin{equation}\label{Friedmann-LQC}
H^2=\frac{8{\pi}G}{3}\rho\left(1-\frac{
\rho}{\rho_c}\right)\,,
\end{equation}
where
\begin{equation}\label{rho-c}
\rho_c=3/(8{\pi}G\alpha_0\gamma^2
l_{_{\rm Pl}}^2)\,,
\end{equation}
where $\gamma$ is the Barbero-Immirzi parameter which should be
fixed by black hole entropy calculations \cite{BI} and $\alpha_0
=4\sqrt{3}\pi\gamma$ is a numerical parameter. In relation
(\ref{Friedmann-LQC}), $\rho\leq \rho_c$ and therefore
$H\leq(\frac{2\pi{ G}}{3}\rho_c)^{1/2}$. Now, with looking at the
result (\ref{Internal-Energy-Bound}) we may modify the effective
Friedmann equation (\ref{Friedmann-LQC}) by following the
identification of the energy density bounds,
\begin{equation}\label{rho}
\rho\leq\,\rho_{\max}=\rho_c\,.
\end{equation}
This identification shows the relevant correspondence between DSR
theory, on one side and LQC on the other side. Relation (\ref{rho})
also determines the numerical value of the entropy bound
(\ref{Entropy-Bound}). The natural identification (\ref{rho}) leads
to the upper bound for the Hubble parameter,
\begin{equation}\label{Hubble}
H\leq\,H_{\max}=\left(\frac{2\pi{G}}{3}
\rho_{\max}\right)^{\frac{1}{2}},
\end{equation}
through the effective Friedmann equation (\ref{Friedmann-LQC}). Note
that the bound for the energy density (or Hubble parameter) in LQC
is obtained from the holonomy-flux algebra through the quantization
of the flat FRW geometry by the method of loop quantum gravity while
bound (\ref{rho-bound}) is obtained in a very different manner {\it
i.e.} by statistical considerations of the particles in the early
Universe in the DSR framework. However, these two different pictures
match each other in a fascinating manner. This result also confirms
that DSR can prepare a suitable framework for the (semiclassical)
flat limit of quantum gravity.

It should also be noted that the scale factor takes a minimum
nonzero value in the context of LQC. This feature of LQC can also be
realized from our setup through the well-known adiabatic condition
for the universe (see also Ref. \cite{SR-NC}),
\begin{equation}\label{adiabatic}
Sa^3=\mbox{cons.}\,,
\end{equation}
where $S$ is the entropy of the radiation dominated universe. The
existence of the entropy bound (\ref{Entropy-Bound}) implies that
there is a nonzero minimum value for the scale factor as
\begin{equation}\label{a-min}
a\geq\,a_{\min}=\left(\frac{\mbox{cons.}}{
S_{\max}}\right)^{\frac{1}{3}}\,.
\end{equation}
Using relation (\ref{Entropy-Bound}) in the above relation and also
applying the fact that $N/V$ is constant for $N,V\rightarrow\infty$,
as one usually assumed in standard statistical mechanics, one can
show that $a_{\min}\sim{l_{_{\rm Pl}}}$. Therefore, the consistency
between the statistical mechanics in the AdS-Cosm DSR model on the
one hand and the results of loop quantum cosmology on the other hand
is completed: The energy density and Hubble parameter approach the
maximum values (\ref{rho}) and (\ref{Hubble}) when the scale factor
approaches the minimum nonzero value (\ref{a-min}). Thus, the
singularity resolution in radiation dominated universe is completely
understood from both the geometrical (LQC) and thermodynamical
(statistical mechanics in DSR) sides.

In summary, the standard thermodynamical results of a radiation
dominated universe match the classical (usual) Friedmann equations
and cannot support the modified Friedmann equation
(\ref{Friedmann-LQC}). On the other hand, the statistical mechanics
based on the AdS-Cosm DSR model matches the effective Friedmann
equation (\ref{Friedmann-LQC}) and cannot match the standard
Friedmann equation (\ref{Friedmann}). However, it is important to
note that although the LQC geometry and the DSR statistical
mechanics are qualitatively consistent, the geometry and matter
parts were not obtained from a unique setup. One then attempts to
explore a bridge between these two setups which seem to be
mathematically and conceptually very different. Nevertheless, any
DSR theory with modified dispersion relation leads to the
modification of geometry such that the spacetime metric becomes
energy dependent \cite{Rainbow}. In this respect, it seems possible
to reobtain the LQC geometry from a DSR theory (or maybe a class of
them) in the context of gravity's rainbow. We are going to study
such a setup for the early radiation dominated universe in the next
research program.

While there is not a clear physical reason to prefer one DSR theory
over the other, thermostatistical consideration suggests that the
theories with a finite total number of microstates such as AdS-Cosm
are more admissible. It should be noted that the AdS-Cosm model
breaks the Lorentz invariance while dS-Cosm and AdS-Stat do not.
Although it is not clear that the Lorentz symmetry will be broken or
deformed at the UV regime, it is interesting to study a DSR theory
which preserves the Lorentz invariance and also supports the
existence of finite total number of microstates. Such a DSR theory
is relevant from both the kinematical and thermodynamical point of
views.

\section{Summary and Conclusions}
Existence of a minimum length scale, below which no other length
scales can be probed, is the main feature of quantum gravity
candidates such as string theory and loop quantum gravity. Although
a complete theory of quantum gravity is not yet formulated, it is
natural to expect that a nongravitational theory which supports the
existence of a minimal length scale arises at the flat limit (weak
gravity limit but high energy regime) of the ultimate quantum theory
of gravity. The DSR theories are then investigated in order to take
into account a minimal observer-independent length scale in special
relativity. These theories are formulated on the dS and AdS momentum
spaces. There are various kinds of DSR theories which can be
realized from the different coordinatization of these curved
momentum spaces. Since the topology of the dS and AdS spaces are
${\mathbf R}\times{{\mathbf S}^3}$ and ${{\mathbf
S}^1}\times{\mathbf R}^3$ respectively, a maximal momentum and
maximal energy naturally arise in dS and AdS momentum spaces,
respectively, by demanding an isotropic (varying) speed of light. In
order to study the associated statistical mechanics, we first
introduced a natural measure on the extended phase space. In light
of the disintegration theorem, we obtained the natural invariant
measure on the physical phase space (the space of the distinct and
physically relevant microstates) by restricting the natural measure
of the extended phase space to the constraint hypersurface and then
fixing the gauge transformation that is generated by the constraint.
By having this invariant measure, one can easily study the
thermostatistics of any DSR theory in any ensemble. Without
attribution to any ensemble density, and quite generally, we have
studied the general statistical properties of four DSR models:
dS-Cosm, dS-Stat, AdS-Cosm, and AdS-Stat. Applying the setup to the
statistical mechanics of the early radiation dominated universe we
have shown that the total number of microstates for the AdS-Cosm
model is finite. We conjecture that this result emerges for two
reasons: having compact energy space and dimensional reduction at
the UV regime. The AdS-Cosm model is the only model that has both of
these properties. We then calculated the corresponding entropy and
internal energy bounds in this model, and we have explored the
cosmological implications of these results. We found that the
geometry of the standard Friedmann equations is no longer applicable
to respect these results since they cannot support the existence of
an upper bound for the energy density of the radiation dominated
universe. We have shown that the AdS-Cosm DSR model respects the
geometry of effective Friedmann equations that arise from the
context of loop quantum cosmology. The existence of a minimum
nonzero scale factor that arises in loop quantum cosmology can also
be understood by means of the resultant entropy bound in the DSR
setup through the adiabatic condition for the universe. Finally, it
seems that the DSR theories which predict a finite total number of
microstates, such as the AdS-Cosm model, are more relevant from the
thermostatistical point of view, and they can be considered as good
candidates for the flat limit of the quantum gravity proposal.

{\bf Acknowledgement}\\
We would like to thank the referees for very insightful
comments which greatly improved the quality of the paper.


\begin{thebibliography}{1}
\bibitem{unit} We work in units $\hbar
=c=k_{_B}=1$, where $\hbar$, $c$,
and $k_{_B}$ are the Planck constant,
speed of light in vacuum, and
Boltzmann constant respectively.
\bibitem{convention} The numerical
factor $4\pi$ is inserted for convenience to achieve the standard
results of statistical mechanics in low energy regime.
\bibitem{IR-TSR} In the context of
triply special relativity, the measure
of position spacetime may be changed
even if one neglects gravity \cite{TSR}.
The curvature of spacetime in such
non-gravitational theories will be due
to an ivariant IR scale and the
cosmological constant may be
interepreted as a universal IR scale
without purely gravitational origin.
\bibitem{DR-NC} The same result is
obtained in Ref. \cite{DR-NC2} from
thermostatistical point of view. It is
shown that the dimensional reduction
in the space of momenta in high
temperature (UV) regime can be
interpreted as the dimensional
reduction in space through invoking to
the equipartition theorem of energy.
\end{thebibliography}
\end{document}